\documentstyle{article}
\date{\today}

\begin{document}
\title{The $\Omega$ and $\Sigma^0\Lambda$ transition magnetic moment 
in QCD Sum Rules}

\author{{Shi-lin Zhu,$^{1,2}$ W-Y. P. Hwang,$^3$ and Ze-sen Yang$^1$}\\
{$^1$Department of Physics, Peking University, Beijing, 100871, China}\\
{$^2$Institute of Theoretical Physics, Academia Sinica}\\
{P.O.Box 2735, Beijing 100080, China}\\
{$^3$Department of Physics, National Taiwan University, Taipei,
Taiwan 10764}
}
\maketitle

\begin{center}
\begin{minipage}{120mm}
\vskip 0.6in
\begin{center}{\bf Abstract}\end{center}
{The method of QCD sum rules in the presence of the external electromagnetic 
$F_{\mu\nu}$ field is used to calculate the $\Omega$ magnetic moment $\mu_{\Omega}$
and the $\Sigma^0\Lambda$ transition 
magnetic moment $\mu_{\Sigma^0\Lambda}$, 
with the susceptibilities obtained previously from the study of octet baryon  
magnetic moments. The results $\mu_{\Omega}=-1.92\mu_N$ 
and $\mu_{\Sigma^0\Lambda}=1.5 \mu_N $ are in good agreement  
with the recent experimental data.\\                          
\\
{\large Keywords: magnetic moment, QCD sum rules}\\
\\
{\large PACS: 13.40.Em, 14.20.Jn, 12.38.Lg}\\
}
\end{minipage}
\end{center}

\large

\vskip 1.5cm
\par 
According to the quark model the $\Omega^-$ is composed 
of three strange quarks with parallel spins.  This state is particularly 
interesting because of the large symmetry breaking that can not be 
accomodated consistently in the naive quark model. 
In Ref. \cite{FRANK}, it has been shown that nonstatic 
baryon-dependent magnetic effects are large. 
In this paper we investigate these symmetry breaking effects and evaluate the 
$\Omega$ moment in the QCD sum rule \cite{SVZ}. 
The $\Omega$ moment has been studied in literaures \cite{3} and various 
theoretical results range from $-1.3 \mu_N$ to $-2.7\mu_N$. 
We find our result is in good agreement with the recent data 
\cite{WALLACE,DATA}.
\par
In a typical hadronic scale the quantum chromodynamics (QCD) is highly 
non-perturbative which makes a direct analytical first-principle calculation
impossible. In this work we adopt the method in the presence of an    
external electromagnetic field \cite{IOFFE,Balit} to calculate the $\Omega$ 
and the $\Sigma^0\Lambda$ transition magnetic moment $\mu_{\Sigma^0\Lambda}$. 
The important aspect of this investigation 
is that the various susceptibilities in this calculation have 
already been determined from previous studies of octet baryon magnetic moments 
\cite{IOFFE,Balit,Kogan,Chiu},  
and as a result our calculation is parameter-free. 

\par
In the method of QCD sum rules \cite{IOFFE,Balit}, the 
two-point correlation function $\Pi(p)$ in the presence of an external 
electromagnetic field $F_{\alpha\beta}$ is written as:
\begin{equation}
\begin{array}{ll}
\Pi_{\Omega} (p)=& i\int d^4 x \langle 0|T \{ \eta_{\mu}(x),{\overline  \eta}^{\mu}(0)\} 
|0\rangle_{F_{\alpha\beta}} e^{ip\cdot x}  \\
&=\Pi_0(p) + \Pi_1 (p) (\sigma\cdot F {\hat p} +{\hat p}\sigma\cdot F) + \cdots \, ,
\end{array}
\end{equation} 
\begin{equation}                                                       
\begin{array}{ll}                               
\Pi_{\Sigma\Lambda} (p)=& i\int d^4 x \langle 0|T \{ \eta_{\Sigma^0}(x),{\overline  \eta}_{\Lambda}(0)\} 
|0\rangle_{F_{\alpha\beta}} e^{ip\cdot x}  \\
&=\Pi_2 (p) (\sigma\cdot F {\hat p} +{\hat p}\sigma\cdot F) + \cdots \, 
\end{array}
\end{equation}
where $\Pi_0(p)$ is the polarization operator without the external field 
$F_{\alpha\beta}$. 
The $\eta_{\mu}$, $\eta_{\Sigma^0}$ and $\eta_{\Lambda}$ 
are the currents with $\Omega$, $\Sigma^0$ and $\Lambda$ quantum numbers
\begin{equation}
\eta_{\mu}(x) =\epsilon^{abc} [{s^a}^T (x) C\gamma_{\mu} s^b(x) ]  s^c(x) \, ,
\end{equation}
\begin{equation}
\eta_{\Sigma^0}(x) =\epsilon^{abc} {1\over \sqrt{2}} \{
[{u^a}^T (x) C\gamma_{\mu} d^b(x) ]\gamma_5 \gamma^{\mu}  s^c(x) 
+[{d^a}^T (x) C\gamma_{\mu} u^b(x) ]\gamma_5 \gamma^{\mu}  s^c(x) \}
\, ,
\end{equation}
\begin{equation}
\eta_{\Lambda}(x) =\epsilon^{abc} {\sqrt{2\over 3}} \{
[{u^a}^T (x) C\gamma_{\mu} s^b(x) ]\gamma_5 \gamma^{\mu}  d^c(x) 
-[{d^a}^T (x) C\gamma_{\mu} s^b(x) ]\gamma_5 \gamma^{\mu}  u^c(x) \}
\, ,
\end{equation}
where $u^a(x)$, $T$ and $C$ are the quark field, the transpose and the 
charge conjugate operators. $a$, $b$, $c$ is the color indices. 
The interpolating currents couples to the baryon states with the 
overlap amplititude $\lambda$. 
\begin{equation}
\langle 0| \eta_{\mu}(0) |\Omega \rangle =\lambda_{\Omega} \nu_{\mu}(p) \, ,
\end{equation}
\begin{equation}
\langle 0| \eta_{\Sigma^0}(0) |\Sigma \rangle =\lambda_{\Sigma} \nu_{\Sigma} (p) \, ,
\end{equation}
\begin{equation}
\langle 0| \eta_{\Lambda}(0) |\Lambda \rangle =\lambda_{\Lambda} \nu_{\Lambda} (p) \, ,
\end{equation}
where the $\nu_\mu$ is a vectorial spinor and satisfies 
$ ({\hat p} -m_{\Omega}) \nu_\mu =0$, 
${\bar \nu}_\mu \nu_\mu =-2m_{\Omega}$, and 
$\gamma_\mu \nu^\mu =p_\mu \nu^\mu =0$ in the Rarita-Schwinger formalism. 
The $\nu (p)$ is a Dirac spinor. 
\par
On the hadronic level the correlators $\Pi_1 (p)$ and $\Pi_2 (p)$  
are expressed in terms of the chirality-odd 
tensor structure $(\sigma\cdot F {\hat p} 
+{\hat p}\sigma\cdot F)$: 
\begin{equation} 
\label{a}
\Pi_1 (p)= -{1 \over 4} \mu_{\Omega} {\lambda_{\Omega}^2 \over (p^2- m^2_\Omega )^2}
\{ {10\over 9} + {4\over 9 m^2_{\Omega} } (p^2- m^2_{\Omega} )^2 \}
(\sigma\cdot F {\hat p} +{\hat p}\sigma\cdot F) +\cdots \, , 
\end{equation}
\begin{equation}
\Pi_2 (p)= -{1 \over 4} \mu_{\Sigma^0\Lambda} 
{\lambda_{\Sigma} \lambda_{\Lambda} \over (p^2- {\bar m}^2)^2}
(\sigma\cdot F {\hat p} +{\hat p}\sigma\cdot F) +\cdots \, .
\end{equation}
Where $\{ \cdots \}$ is $1$ for the nucleon magnetic moment. 
The deviation from unity in (\ref{a}) is due to 
the Rarita-Schwinger formalism for the spin ${3\over 2}$ field,  
of which the detail may be found in Ref. \cite{TAKHA}. 
${\bar m}= {m_{\Sigma^0} +m_{\Lambda} \over 2}$. We treat the $\Sigma$ and 
$\Lambda$ mass as degenerate due to their small mass difference since we  
never come across the poles and always work in the virtulity $p^2 < 0$. 
We denote the continuum and non-diagonal transition 
contributions simply by ellipse.
\par
The external field $F_{\mu\nu}$ may induce changes in the physical vacuum and 
modify the propagation of quarks. Up to dimension six ($d\leq 6$), we 
introduce three induced condensates:
\begin{equation} 
\begin{array}{c}
\langle 0 | {\overline  q} \sigma_{\mu\nu} q |0 \rangle_{F_{\mu\nu}} = e_q  \chi
F_{\mu\nu} \langle 0 | {\overline  q}  q |0 \rangle \, , \\ 
g_s \langle 0 | {\overline  q} {{\lambda^n}\over {2}}G^n_{\mu\nu} q |0 \rangle_{F_{\mu\nu}}
= e_q  \kappa F_{\mu\nu} 
\langle 0 | {\overline  q}  q |0 \rangle \, , \\
g_s \epsilon^{\mu\nu\lambda\sigma} 
\langle 0 | {\overline  q} \gamma_5 {{\lambda^n}\over {2}} G^n_{\lambda\sigma} q |0 \rangle_{F_{\mu\nu}}
= i e_q  \xi F^{\mu\nu} 
\langle 0 | {\overline  q}  q |0 \rangle \, ,
\end{array}
\end{equation}
where $q$ refers to $u$, $d$ and $s$ quark, $e_q$ is the charge. 
The $\chi$, $\kappa$ and $\xi$ in Eq. (5) are the quark condensate 
susceptibilities and their values have been the subject of various 
studies. Ioffe and Smilga \cite{IOFFE} found 
$\chi \approx -8\,\mbox{GeV}^{-2}$ with $\kappa =0$, $\xi=0$ in order  
to have $\mu_p =3.0 \mu_N$ and $\mu_n=-2.0\mu_N$ $(\pm 10\%)$.  
Balitsky and Yung \cite{Balit} estimated 
\begin{equation}
\begin{array}{lll}
\chi=-3.3\,\mbox{GeV}^{-2},& \kappa=0.22,& \xi=-0.44 \, .
\end{array}
\end{equation}
with the one-pole approximation. 
Belyaev and Kogan \cite{Kogan} extended the calculation and 
obtained an improved estimate $\chi =-5.7\,\mbox{GeV}^{-2}$ using the two-pole 
approximation. Chiu et al \cite{Chiu} also estimated the susceptibilities 
with two-pole model and obtained
\begin{equation}
\begin{array}{lll}
\chi=-4.4\,\mbox{GeV}^{-2},& \kappa=0.4,& \xi=-0.8 \, .
\end{array}
\end{equation}
The values of these susceptibilities are consistent 
with one another except that the earliest result $\chi=-8\,\mbox{GeV}^{-2}$ 
in \cite{IOFFE}, is considerably larger (in magnitude) due to their  
neglect of $\kappa$, $\xi$ in the fitting procedure. In what 
follows, we shall adopt the condensate parameters $\chi=-4.5\,
\mbox{GeV}^{-2}$,  
$\kappa =0.4$, $\xi = -0.8$  which represent the 
average in the last three analyses. 
\par
The correlation functions $\Pi_1 (p)$ and $\Pi_2 (p)$ at the quark level are 
\begin{equation}
\begin{array}{ll}
\label{b}
\langle 0| T\eta_{\mu}(x) {\bar \eta}^{\mu}(0) |0\rangle_F =
-2 {\rm i} \epsilon^{abc}
\epsilon^{a^{\prime} b^{\prime} c^{\prime}}(& {\bf Tr} \{ S^{b b^{\prime}}(x)
\gamma_{\mu} C [S^{a a^{\prime}}(x) ]^T C \gamma_{\mu} \}\,
S^{c c^{\prime}}(x) \\
&
+2  S^{b b^{\prime}}(x)
\gamma_{\mu} C [S^{a a^{\prime}}(x) ]^T C \gamma_{\mu} S^{c c^{\prime}}(x) \, \, ) \, ,
\end{array}
\end{equation}
\begin{equation}
\begin{array}{ll}
\label{c}
\langle 0| T\eta_{\Sigma^0}(x) {\bar \eta}^{\Lambda}(0) |0\rangle_F =
-{2\over \sqrt{3}} 
{\rm i} \epsilon^{abc}                  
\epsilon^{a^{\prime} b^{\prime} c^{\prime}}\{
& \gamma_5 \gamma^{\mu} S_s^{a a^{\prime}}(x)
\gamma_{\nu} C [S_u^{b b^{\prime}}(x) ]^T C \gamma_{\mu} 
S_d^{c c^{\prime}}(x) \gamma^{\nu}\gamma_5 \\
&
- \gamma_5 \gamma^{\mu} S_s^{a a^{\prime}}(x)
\gamma_{\nu} C [S_d^{b b^{\prime}}(x) ]^T C \gamma_{\mu} 
S_u^{c c^{\prime}}(x) \gamma^{\nu}\gamma_5 
\}
\end{array}
\end{equation}
where $iS^{ab}(x)$ is the quark propagator in the presence of the external 
eletromagnetic field $F_{\mu\nu}$ \cite{IOFFE}. 
We find 
\begin{equation}
\begin{array}{ll}
iS^{ab}_q(p)=
& \delta^{ab}  {{i} \over { {\hat p} -m_q}} \\
&+ {i\over 4} {\lambda_{ab}^n \over 2} g_s G_{\mu\nu}^n {1\over (p^2-m_q^2)^2}
       [\sigma^{\mu\nu} ({\hat p}+m_q) +({\hat p}+m_q) \sigma^{\mu\nu}) ] \\
&+ {i\over 4} e_q \delta^{ab} F_{\mu\nu}  {1\over (p^2-m_q^2)^2}
       [\sigma^{\mu\nu} ({\hat p}+m_q) +({\hat p}+m_q) \sigma^{\mu\nu})  ]  \\
&-\delta^{ab}  { \langle {\overline q}q \rangle \over 12} (2\pi )^4\delta^4(p)\\
&-\delta^{ab}  { \langle g_s {\overline q} \sigma \cdot G q \rangle \over 192} 
  (2\pi )^4 g^{\mu\nu} \partial_\mu \partial_\nu \delta^4(p) \\
&-\delta^{ab}e_q  { \langle {\overline q}q \rangle \over 192} 
  [ \sigma\cdot F g^{\mu\nu} - {1\over 3} \gamma^{\mu} \sigma\cdot F 
  \gamma^{\nu} ] (2\pi )^4 \partial_\mu \partial_\nu \delta^4(p)  \\
&-\delta^{ab} e_q \chi  { \langle {\overline q} q \rangle \over 24} 
  \sigma\cdot F(2\pi )^4\delta^4(p)                               \\
&-\delta^{ab}e_q \kappa  { \langle  {\overline q} q \rangle \over 192} 
  [ \sigma \cdot F g^{\mu\nu}- {1\over 3}\gamma^{\mu}
  \sigma \cdot F \gamma^{\nu} ]
  (2\pi )^4 \partial_\mu \partial_\nu \delta^4(p)                   \\
&+i \delta^{ab}e_q \xi  { \langle  {\overline q} q \rangle \over 768} 
  [ \sigma^{\delta\rho}  g^{\mu\nu}- {1\over 3}\gamma^{\mu} 
  \sigma^{\delta\rho}  \gamma^{\nu} ]\gamma_5 \epsilon_{\alpha\beta\delta\rho} F^{\alpha\beta}
  (2\pi )^4 \partial_\mu \partial_\nu \delta^4(p)                     \\
&+ \cdots,
\end{array}
\end{equation}
in the momentum space with ${\hat p}\equiv p_\mu \gamma^\mu$. 
Here we follow \cite{IOFFE,Balit} 
and do not introduce induced condensates of higher dimensions, while we 
find that the strange quark mass correction is important 
so that we have explicitly kept it in our calculation.
\par
Only the condensates with even dimensions contribute to the structure 
$(\sigma\cdot F {\hat p} +{\hat p}\sigma\cdot F)$. They are $1$, 
$\chi \langle 0| {\bar q} q |0\rangle m_s$, 
$m_s \langle 0| {\bar q} q |0\rangle$, 
$\langle 0|g_s^2G^n_{\alpha\beta}G^n_{\alpha\beta}|0\rangle $, 
$\chi \langle 0| {\bar q} q |0\rangle^2$, 
$\chi \langle 0| {\bar q} q |0\rangle \langle 0|g_s {\bar q} \sigma\cdot G q |0\rangle$, 
$\kappa\langle 0| {\bar q} q |0\rangle^2$, 
and 
$\xi \langle 0| {\bar q} q |0\rangle^2$ 
up to dimension eight. The up and down quark is treated as massless. 
The gluon condensate
$\langle 0|g_s^2G^n_{\alpha\beta}G^n_{\alpha\beta}|0\rangle $ always appears 
with a small numerical factor ${1\over (2 \pi)^4}$ through the two-loop 
integration \cite{IOFFE}. Its contribution was found to be negligible                                                         
through the direct calculation \cite{Chiu}. 
Following Refs. \cite{SVZ,IOFFE,Balit} 
the four quark condensate 
$\langle 0| {\bar q} \Gamma_1 q {\bar q} \Gamma_2 q|0\rangle$ is treated 
by the factorization approximation as 
the susceptibilities are estimated under the the vacuum dominance hypothesis.
The calculation is staightforward by substituting the quark propagator
into equation (\ref{a}) and (\ref{b}). Here we present the final result after 
Borel transformation. 
\begin{equation}
\label{z1}
\begin{array}{ll}
{9\over 4} e_s 
& \{
 M^6_B E_2(y_1) L^{\frac{4}{27}} 
 -{6\over 5}\chi m_s a_s M_B^4 E_1(y_1) L^{-\frac{4}{9}}
 +{6\over 5} m_s a_s M_B^2 E_0(y_1) L^{\frac{4}{27}}    \\
& -{4\over 5}\chi a_s^2 M_B^2 E_0(y_1) L^{\frac{4}{9}}
  +a_s^2 L^{\frac{28}{27}} 
  [{14\over 15} +{\chi m_0^2 \over 30}L^{-\frac{10}{9}}
  -{22\over 15}\kappa -{4\over 15}\xi ] \} \\
&= (2\pi )^4 \lambda_{\Omega}^2 e^{-\frac{m_{\Omega}^2}{M_B^2}}
 \mu_{\Omega} (1+A_1 M_B^2) \, ,
\end{array}
\end{equation}
\begin{equation}
\label{z2}
\begin{array}{ll}
{1\over \sqrt{3}}(e_u - e_d)
& \{
 M^6_B E_2(y_2) L^{-\frac{4}{9}} 
 -\chi m_s a M_B^4 E_1(y_2) L^{-\frac{28}{27}}
 +{4\over 3} m_s a M_B^2 E_0(y_2) L^{-\frac{4}{9}}   \\ 
& - m_s a_s M_B^2 E_0(y_2) L^{-\frac{4}{9}}  
  -{2\over 3}\chi a a_s M_B^2 E_0(y_2) L^{-\frac{4}{27}}
  +{1\over 9}a a_s L^{\frac{4}{9}} 
  [ 4 +\kappa -2\xi
  +{3 \over 4}\chi m_0^2L^{-\frac{10}{9}} ] 
  \} \\
&= (2\pi )^4 \lambda_{\Sigma} \lambda_{\Lambda}
  e^{-\frac{{\bar m}^2}{M_B^2}}
 \mu_{\Sigma^0\Lambda} (1+A_2 M_B^2)
\end{array}
\end{equation}
where $m_{\Omega}=1.672\mbox{GeV}$, ${\bar m} =1.15\mbox{GeV}$, 
$m_s =150$MeV, $y_1=\frac{W_1^2}{M_B^2}$ and $y_2=\frac{W_2^2}{M_B^2}$.  
$E_n (y)=1-e^{-y} \sum\limits_{k=0}^{n} \frac{1}{k!} y^k $ 
are the factors used to subtract the continuum contribution \cite{IOFFE}. 
$W_1^2=5.0$GeV$^2$ and $W_2^2=3.4$GeV$^2$ 
are the continuum thresholds which are 
determined together with the overlap amplititudes 
$(2\pi )^4\lambda_{\Omega}^2 =5.56$GeV$^6$,  
$(2\pi )^4\lambda_{\Sigma}^2 =1.88$GeV$^6$ and   
$(2\pi )^4\lambda_{\Lambda}^2 =1.64$GeV$^6$  
from the $\Omega$, $\Sigma$ and $\Lambda$ 
mass sum rules \cite{REINDERS,HW}. 
We adopt the ``standard'' values for the various condensates 
$a=-(2\pi )^2 \langle 0 | {\overline  u}  u |0 \rangle 
=0.55   \mbox{GeV}^3 $, 
$a_s=-(2\pi )^2 \langle 0 | {\overline  s}  s |0 \rangle 
=0.55  \times 0.8 \mbox{GeV}^3 $, 
$a m_0^2 =(2\pi )^2 g_s \langle 0 | {\overline  u}\sigma\cdot G  u |0 \rangle $, 
$m_0^2 =0.8\mbox{GeV}^2$.
$L=\frac{\ln (\frac{M_B}{\Lambda_{\mbox{QCD}}})}{\ln (\frac{\mu}{\Lambda_{\mbox{QCD}}})}$, 
$\Lambda_{\mbox{QCD}}$ is the QCD parameter, $\Lambda_{\mbox{QCD}}= 100$MeV, 
$\mu = 0.5$GeV is the normalization point to which the used values of 
condensates are referred. 
\par
We further improve the numerical analysis by taking into account of the 
renormalization group evolutions of the sum rules (\ref{z1}) and 
(\ref{z2}) through 
the anomalous dimensions of the various condensates and currents. 
$A_1$ and $A_2$ are constants to be determined from the sum rule, 
which arise from the nondiagonal transitions  
$\Omega \gamma\to \Omega^{\ast}$, 
$\Sigma^{\ast} \to\Lambda\gamma$ or
$\Sigma \to \Lambda^{\ast} \gamma$ \cite{IOFFE}. 
The working intervals of the Borel mass $M_B^2$ 
for the sum rule (\ref{z1}) and (\ref{z2}) 
are $2.0 \mbox{GeV}^2 \leq M_B^2 \leq 4.0 \mbox{GeV}^2$ 
and $1.3 \mbox{GeV}^2 \leq M_B^2 \leq 3.0 \mbox{GeV}^2$ 
respectively where both the 
continuum contribution and power corrections are controllable.
Moving the factor $(2 \pi )^4 \lambda_{\Omega}^2 e^{- {m^2\over M_B^2}}$
and $(2 \pi )^4 \lambda_{\Sigma} \lambda_{\Lambda}
e^{- {{\bar m}^2\over M_B^2}}$ on the right hand side to the left and 
fitting the new sum rules with a straight line approximation 
we may extract the $\mu_{\Omega}$ and $\mu_{\Sigma\Lambda}$.  
We show the new sum rules as a function of the Borel mass in Fig. 1.  
The new sum rules are almost stable and independent of 
$M_B^2$, which implies that the nondiagonal tansition  
contributions are small. The sum rules are insensitive to the 
susceptibilities $\kappa$ and $\xi$ due to their small values. 
Their contributions are less than $5\%$. 
The dependence on $\chi$ is shown in Fig. 1. 
When $\chi$ varies from $-4.5$GeV$^{-2}$ to $-3.5$GeV$^{-2}$ or 
to $-5.5$GeV$^{-2}$, the sum rules change within $10\%$. 
The correction from the strange quark mass 
is important and contributes about $20\%$ to both of the sum rules.  
The $SU(3)_f$ flavour symmetry breaking 
parameter $\gamma = {\langle {\bar s} s \rangle \over 
\langle {\bar u} u \rangle }$ need to take the standard 
value of $0.8$ in order to yield a good agreement with 
the experimental data in Eq. (\ref{z1}). The $\Omega$ magnetic moment 
would increase $30\%$ if $\gamma =1$, in 
contradiction with the experimental data. 
Our final results are $\mu_{\Omega}=-3.41 $ in unit of 
${e\over 2m_{\Omega}}$ and $\mu_{\Sigma^0\Lambda}=1.85 $ in unit of 
${e\over 2{\bar m}}$, where ${e\over {2 m_B}}$ is   
a natural unit in QCD sum rule analysis. 
In unit of nuclear magneton $\mu_{\Omega}=-1.92 \mu_N$ and  
$\mu_{\Sigma^0\Lambda}=1.5 \mu_N$, 
in good agreement with the recent experimental data. 
\par
Baryon magnetic moments are important physical observables as masses. 
The method of QCD sum rules in the presence of an external 
electromagnetic field was successfully employed to calculate 
the octet baryon magnetic moments.  
The results are in reasonable agreement with the experimental data. 
In this work we have extended the same method to calculate 
the magnetic moment of the long-lived decuplet member, the $\Omega$, 
and $\Sigma^0\Lambda$ transition magnetic moment simultaneously, 
which may serve both as a consistency check 
of the various susceptibilities and a check of the method of 
the external field itself. Our results are in good agreement                                                               
with the recent experimental data.
\par
This work is supported in part by the Postdoctoral Science Foundation
of China and the National Natural Science Foundation
of China. It is also supported in part by the National Science Council of 
R.O.C. (Taiwan) under the grant NSC84-2112-M002-021Y.

\newpage
\begin{center}
{\bf Figure Captions}
\end{center}
\noindent
Fig. 1 (a) The Borel mass dependence of the $\Omega$ magnetic moment. The 
long-dashed, solid and short-dashed 
curve is the QCD sum rule prediction for 
$\chi=-3.5$GeV$^{-2}$, $-4.5$GeV$^{-2}$ and $-5.5$GeV$^{-2}$ respectively
from equation (\ref{z1}) after the 
numerical factor $(2 \pi )^4 \lambda_{\Omega}^2
e^{- {{m_{\Omega}}^2\over M_B^2}}$ is moved to the left hand
side. The dotted line is a straight-line approximation. The intersect with 
Y-axis is the $\Omega$ magnetic moment in unit of ${e\over 2m_{\Omega}}$. The 
Borel mass $M_B^2$ is in unit of GeV$^2$. 
(b) The Borel mass dependence of the $\Sigma\Lambda$ transition 
magnetic moment. Notations are the same as those in (a).


\begin{thebibliography}{99}
\bibitem{FRANK}J. Franklin, Phys. Rev. D {\bf 29}, (1984) 2648.
\bibitem{3}
Y. Tomowaza, Phys. Rev. D {\bf 19} (1979) 1626; \\
T. Das and S. P. Misra, Phys. Lett B {\bf 96} (1980) 165;\\
C. Benard et al., Phys. Rev. Lett. {\bf 49} (1982) 1076;\\
H. J. Lipkin, Nucl. Phys. B {\bf 214} (1982) 136;\\
H. Georgi and A. Monohar, Phys. Lett. B {\bf 132} (1983) 183;\\
V. P. Efrosinin and D. A. Zaikin, Yad. Fiz {\bf 44} (1986) 1053
[Sov. J. Nucl. Phys. {\bf 44} (1986) 681]; \\
R. C. Verma and M. P. Khanna, Phys. Lett. B {\bf 183} (1987) 207;\\
L. Brekke and J. L. Rosner, Comments Nucl. Part. Phys. {\bf 18} (1988) 83;\\
M. Krivoruchenko et al., Phys. Rev. D {41} (1990) 997; \\
J. Kunz and P. J. Mulders, Phys. Rev. D {\bf 41} (1990) 1578.
\bibitem{SVZ}M. A. Shifman, A. I. Vaishtein, and V. I. Zakharov, Nucl. Phys.
B {\bf 147} (1979) 385, 448.
\bibitem{WALLACE}N. B. Wallace et al., Phys. Rev. Lett. {\bf 74} (1995) 3732.
\bibitem{DATA}Particle Data Group, Phys. Rev. D {\bf 54} (1996) 55.
\bibitem{IOFFE}B. L. Ioffe and A. V. Smilga, Nucl. Phys. B {\bf 232} (1984) 109.
\bibitem{Balit}I. I. Balitsky and A. V. Yung, Phys. Lett. B {\bf 129} (1983) 328.
\bibitem{Kogan}V. M. Belyaev and Ya. I. Kogan, Yad. Fiz {\bf 40} (1984) 1035.
(Sov. J. Nucl. Phys. {\bf 40} (1984) 659.)
\bibitem{Chiu}C. B. Chiu, J. Pasupathy, and S. J. Wilson, Phys. Rev. D {\bf 33} (1986) 
1961; C. B. Chiu, S. L. Wilson, J. Pasupathy, and J. P. Singh, Phys. Rev. 
D {\bf 36} (1987) 1451, 1553.
\bibitem{TAKHA}Y. Takhashi, {\sl An Introduction to Field Quantization}, 
Pergamon Press, 1969.
\bibitem{REINDERS}L. J. Reinders, H. Rubinstein, and S. Yazaki, Phys. Lett. B 
{\bf 120} (1983) 209.
\bibitem{HW}W-Y. Hwang and K-C. Yang, Phys. Rev. D {\bf 49} (1994) 460.
\end{thebibliography}
\end{document}